\documentstyle[12pt]{article}
\title{ Constraints on Masses of Charged PGBs in Technicolor Model
from  Decay $b\to s\gamma$}
\author{ Cai-Dian  L\"{u}$^{a,b}$ and  Zhenjun Xiao$^{a,c}$\\
{\small a. CCAST(World Laboratory), P.O.Box 8730, Beijing 100080,
P.R.China}\\
{\small b. Institute of Theoretical Physics, Academia Sinica, Beijing 100080,
P.R.China\thanks{Mailing address}}\\
{\small c. Physics Department, Henan Normal University,
Xinxiang, Henan, 453002, P.R.China$^*$}\\ }
\date{}

\begin{document}
\maketitle
\begin{picture}(0,0)(0,0)
\put(310,270){{\large hep-ph/9508380}}
\put(310,250){{\large AS-ITP/95-20}}
\put(310,230){{\large HNU-TH/95-09}}
\put(310,210){{\large May, 1995}}
\end{picture}
\begin{abstract}
In this paper we calculate the contributions to the branching ratio of
$B\to X_s \gamma$ from the charged Pseudo-Goldstone bosons appeared in
one generation Technicolor model. The current $CLEO$ experimental results
can eliminate large part of the parameter space in the
$m(P^\pm) - m(P_8^\pm)$ plane, and specifically, one can put a
strong lower bound
on the masses of color octet charged PGBs $P_8^\pm$:
$m(P^{\pm}_8) > 400\;GeV$ at $90\%C.L$ for free $m(P^{\pm})$.
\end{abstract}
\vspace{1cm}


\newpage
\section{Introduction}

Recently the CLEO collaboration has observed\cite{cleo1} the
exclusive radiative decay $B \rightarrow K^* \gamma $ with a
branching fraction
of $BR(B\rightarrow K^* \gamma)=(4.5 \pm1.0 \pm0.9)\times10^{-5}$.
The inclusive $b\to s \gamma$ branching ratio measured by CLEO\cite{cleo2}
is:
\begin{eqnarray}
BR(B\to X_s \gamma ) = (2.32\pm 0.57 \pm 0.35) \times 10^{-4}.
\end{eqnarray}
The newest upper and lower limits of this decay branching ratio are
\begin{eqnarray}
1.0\times 10^{-4} < BR(B \to X_s \gamma) < 4.2\times 10^{-4},
\ \ at \ \ 95\%C.L..
\end{eqnarray}
As a loop-induced flavor changing neutral current(FCNC)
process the inclusive  decay(at quark level) $b\to s \gamma$ is
in particular sensitive to contributions
from those new physics beyond the Standard Model(SM)\cite{Hew1}.
There is a vast interest in this decay.

The decay $b\to s \gamma$ and its large leading log QCD corrections
have been evaluated in the  SM by several groups
\cite{Ciu}. The reliability of the calculations of this
decay is improving as partial calculations of the next-to-leading
logarithmic QCD corrections to the effective Hamiltonian\cite{AJB,lcd}.

On the other hand the discovery of the top quark and the measurement
of its mass (in this paper we use the
weighted average $m_t=180\pm 12\; GeV$ from the announced
results of $m_t$ by CDF and D0\cite{cdfd0} wherever possible)
at FERMILAB basically eliminated a source of uncertainties for the
calculation of the decay $b\to s \gamma$ in the SM and in beyond theories.
The great progress in theoretical studies and in experiments achieved
recently encourage us to do more investigations about this decay in
Technicolor theories.

In this paper, we estimate the possible contributions
to the decay  $b\to s \gamma$
from the exchange of the charged Pseudo-Goldstone bosons which will
appear  in no-minimal Technicolor models, such as the Farhi-Susskind
one-generation Technicolor model (OGTM) \cite{Farhi}.
We know that the experimental data seems
disfavor the OGTM which generally tend to predict $S$ parameter
large and positive \cite{Peskin92}. Why we here
still choose it to do the calculations?  The reasons are the following:

(1) At first, presence of the Pseudo-Goldstone bosons in the particle spectrum
is a common feature of those non-minimal TC models with ordinary or
novel ETC sectors, no matter the specific differences of structures
between those models.
The gauge couplings of the PGBs are determined by their quantum numbers,
while the Yukawa couplings of PGBs to ordinary fermions
are generally proportional to fermion masses
for many TC/ETC models.
Among the non-minimal TC models,
the OGTM \cite{Farhi} is the simplest and most frequently studied
model. Many relevant works
\cite{King95} have been done since the late 1970's.
One can use those existed results directly in  further investigations.

(2) On the other hand, the constraints on the $S$ parameter could be relaxed
considerably by introducing three additional parameters
$(V, W, X)$ \cite{Burgess94a}.
A global fit to the data in which
all six oblique parameters S through X are allowed to vary
simultaneously gives the one standard deviation bound on $S$:
$ S \sim -0.93 \pm 1.7$ \cite{Burgess94b}.
This fact means that the constraint on the OGTM
from the parameter $S$ could be considerably weakened if we
consider the effects from light technifermions and light
PGBs \cite{Evans94}.

In this paper, we estimate the possible
contributions to the rare decay $b\to s \gamma$ from the charged PGBs in the
framework
of the OGTM. At least, one can
regard our results as an  estimation for the ``correct'' output of
the future ``realistic'' TC models.

This paper is organized as the following: In Section 2, we  present the
basic ingredients of the OGTM and then calculate the PGB contributions
to $b\to s \gamma $ decay, together with the full leading log QCD
corrections. In Section 3, we obtain the branching ratios of this decay,
and derive out the
constraints on masses of charged PGBs by phenomenological analysis.
The conclusions are also included in this
section.

\section{Charged PGBs and QCD Corrections to $b\to s \gamma$}

In the OGTM \cite{Farhi}, when the technifermion condensate
$<\overline{T}T>\neq 0$
was formed, the global flavor symmetry will break as follows:
$SU(8)_L \times SU(8)_R \rightarrow SU(8)_{L+R}$.
Consequently, 63 (Pseudo)-Goldstone bosons will
be produced from this breaking.
When all other interactions but the Technicolor are turned off, these 63
Goldstone bosons are exactly massless. Three of them are eaten by the
$W^{\pm}$ and $Z^0$ gauge bosons. The others acquire masses
when one turned on the
gauge interactions, and therefore they are Pseudo-Goldstone Bosons(PGBs).

According to previous studies, the phenomenology of those color-singlet
charged PGBs in the OGTM is very similar with that of the elementary
charged Higgs bosons $H^{\pm }$ of Type-I Two-Higgs-Doublet
Model(2HDM) \cite{Gunion}. And consequently, the contributions to the decay
$b\rightarrow s\gamma$ from the color-singlet
charged PGBs in the OGTM will be very similar with that from
charged Higgs bosons in the 2HDM.
As for the color-octet charged PGBs, the situation is more complicated
because of the involvement of the color interactions.
Other neutral PGB's don't contribute to the rare decay
$b\to s \gamma$.

The gauge couplings of the PGBs are determined by their quantum numbers.
The Yukawa couplings of PGBs to ordinary fermions are induced by ETC
interactions and hence are model dependent. However, these Yukawa couplings
are generally proportional to fermion masses with small differences in the
magnitude of the coefficients for different TC/ETC models.
The relevant couplings needed in our
calculation are directly quoted from refs.\cite{Ellis,Chivukula95,Xiao94}
and summarized in Table \ref{table1},
where the $V_{ud} $ is the corresponding element of Kobayashi-Maskawa
 matrix. For the OGTM, the Goldstone boson decay constant
$F_\pi$ in Table \ref{table1}
should be $F_{\pi}=v/2=123\;GeV$,
in order to ensure the correct masses for the gauge bosons
$Z^0$ and $W^{\pm}$ \cite{King95}.

\begin{table}[htbp]
\caption{ The relevant gauge couplings and Effective Yukawa couplings for
the OGTM. }\label{table1}
\begin{center}
\vspace{.1cm}
\begin{tabular}{|c|c|} \hline
$P^+ P^- \gamma_\mu$  & $-ie(p_+ - p_-)_\mu$ \\ \hline
$P^+_{8a} P^-_{8b} \gamma_\mu$  & $-ie(p_+ - p_-)_\mu  \delta_{ab}$ \\ \hline
$P^+\; u\; d$  & $i\frac{V_{ud}}{2 F_\pi}\sqrt{\frac{2}{3}}
[M_u (1-\gamma_5) - M_d (1 + \gamma_5) ]$ \\ \hline
$P^+_{8a}\; u\; d$  & $i\frac{V_{ud}}{2 F_\pi} 2 \lambda_a
[M_u (1-\gamma_5) - M_d (1 + \gamma_5) ]$ \\ \hline
$P^+_{8a} P^-_{8b} g_{c\mu}$  & $-g f_{abc}(p_a - p_b)_\mu $ \\ \hline
\end{tabular}
\end{center}
\end{table}

In ref.\cite{Randall}, Randall and Sundrum have estimated the contributions to
$b\to s \gamma$ from the exchange of ETC gauge bosons in various
ETC scenarios. In the case of ``traditional'' ETC (just the case which
will be studied here), the dominant contribution to $b\to s \gamma$
occurs when the ETC gauge boson is exchanged between purely left-handed
doublets and when the photon is emitted from the
technifermion line. But the resulted ETC contribution is strongly
suppressed with respect to the SM by a factor of $m_t/(4\pi v)< 0.09$ for
$m_t< 200\;GeV$\cite{Randall}. In short, the ETC contribution to the decay
$b\rightarrow s\gamma$ is small and will be masked by still large
experimental  and theoretical uncertainties. We therefore can
neglect the ETC Contributions to $b\to s \gamma$ at present
phenomenological analysis.

In Fig.1, we draw the relevant Feynman diagrams which contribute to the decay
$b\to s \gamma$, where the half-circle lines represent the W
gauge boson of SM as well as the charged PGBs $P^{\pm}$ and
$P_8^{\pm}$ of OGTM.
In the evaluation we at first integrate
out the top quark and the weak W bosons at $\mu=M_W$ scale,
generating an effective five-quark theory. By using the
renormalization group equation, we run the effective field theory
down to b-quark scale to give the leading log QCD corrections,
then at this scale, we calculate the rate of radiative $b$ decay.

After applying the full QCD equations of motion\cite{eom}, a complete
set of dimension-6 operators relevant for $b\to s \gamma $ decay can be
chosen to be:
\begin{eqnarray}
O_1&=&(\overline{c}_{L\beta} \gamma^{\mu} b_{L\alpha})
	    (\overline{s}_{L\alpha} \gamma_{\mu} c_{L\beta})\;,\\
O_2&=&(\overline{c}_{L\alpha} \gamma^{\mu} b_{L\alpha})
	    (\overline{s}_{L\beta} \gamma_{\mu} c_{L\beta})\;,\\
O_3&=&(\overline{s}_{L\alpha} \gamma^{\mu} b_{L\alpha})
\sum_{q=u,d,s,c,b}(\overline{q}_{L\beta} \gamma_{\mu} q_{L\beta})\;,\\
O_4&=&(\overline{s}_{L\alpha} \gamma^{\mu} b_{L\beta})
\sum_{q=u,d,s,c,b}(\overline{q}_{L\beta} \gamma_{\mu} q_{L\alpha})\;,\\
O_5&=&(\overline{s}_{L\alpha} \gamma^{\mu} b_{L\alpha})
\sum_{q=u,d,s,c,b}(\overline{q}_{R\beta} \gamma_{\mu} q_{R\beta})\;,\\
O_6&=&(\overline{s}_{L\alpha} \gamma^{\mu} b_{L\beta})
\sum_{q=u,d,s,c,b}(\overline{q}_{R\beta} \gamma_{\mu} q_{R\alpha})\;,\\
O_7&=&(e/16\pi^2) m_b \overline{s}_L \sigma^{\mu\nu}
	    b_{R} F_{\mu\nu}\;,\\
O_8&=&(g/16\pi^2) m_b \overline{s}_{L} \sigma^{\mu\nu}
	    T^a b_{R} G_{\mu\nu}^a\;.
\end{eqnarray}

The effective Hamiltonian appears just below the W-scale is given as
\begin{equation}
{\cal H}_{eff} =\frac{4G_F}{\sqrt{2}} V_{tb}V_{ts}^*
	\displaystyle{\sum_{i=1}^{8} }C_i (M_W^-) O_i(M_W^-).
\end{equation}
The coefficients of 8 operators are calculated from diagrams of Fig.1:
\begin{eqnarray}
 C_i(M_W)&=&0, \;\; i=1,3,4,5,6, \;\;\; C_2(M_W)=-1,\\
C_7(M_W)&=& A(\delta) +\frac{B(x)}{3\sqrt{2}G_F F_{\pi}^2 }
+\frac{8 B(y)}{3\sqrt{2}G_F F_{\pi}^2 },
\label{c7}\\
C_8(M_W)&=& C(\delta)
+\frac{D(x)}{3\sqrt{2}G_F F_{\pi}^2 }
+\frac{8 D(y) + E(y)}{3\sqrt{2}G_F F_{\pi}^2 },\label{c8}
\end{eqnarray}
with $\delta=M_W^2/m_t^2$, $x=(m(P^{\pm})/m_t)^2$ and
$y=(m(P^{\pm}_8)/m_t)^2$.
The functions $A$ and $C$ arise from graphs with W boson exchange
are already known contributions from SM; while the functions $B$, $D$,
and $E$ arise from diagrams with color-singlet and color-octet charged
PGBs of OGTM. They are given  by,
\begin{eqnarray}
A(\delta)&=& \frac{ \frac{1}{3} +\frac{5}{24} \delta -\frac{7}{24}
	\delta^2}{(1-\delta)^3}
	+\frac{ \frac{3}{4}\delta -\frac{1}{2}\delta^2}{(1-\delta)^4}
\log[\delta] \\
B(y)& =&  \frac{ -\frac{11}{36}
	+\frac{53}{72}y -\frac{25}{72}y^2}{(1-y)^3}
	+\frac{ -\frac{1}{4}y +\frac{2}{3}y^2 -\frac{1}{3}y^3}
	{(1-y)^4}\log[y],\\
C(\delta)&=& \frac{\frac{1}{8} -\frac{5}{8} \delta-\frac{1}{4}
	\delta^2}{(1-\delta)^3}
	-\frac{ \frac{3}{4}\delta^2}{(1-\delta)^4} \log[\delta] \\
D(y)& =& \frac{ -\frac{5}{24}
	+\frac{19}{24}y -\frac{5}{6}y^2}{(1-y)^3}
	+\frac{ \frac{1}{4}y^2 -\frac{1}{2}y^3}{(1-y)^4}
	\log[y],\\
E(y) & =& \frac{ \frac{3}{2}-\frac{15}{8}y -\frac{5}{8}y^2 }{(1-y)^3}
	+\frac{\frac{9}{4}y -\frac{9}{2}y^2}{(1-y)^4 }\log[y].
\end{eqnarray}
It is shown from these expressions that, for $\delta < 1$, $x,y >> 1$,
the OGTM contribution $B$, $D$ and $E$ have always a relative minus
sign with the SM contribution $A$ and $C$. As a result, the OGTM
contribution always destructively interferes with the SM contribution.
This can also be seen from the numerical results and discussion
in the next section.

The running of the coefficients of operators from $\mu=M_W$ to $\mu=m_b$
was well described in refs.\cite{Ciu}. After  renormalization
group running we have the QCD corrected
coefficients of operators at $\mu=m_b$ scale.
\begin{equation}
C_7^{eff}(m_b) = \eta^{16/23}C_7(M_W) +\frac{8}{3}
( \eta^{14/23}-\eta^{16/23} ) C_8(M_W)
+C_2(M_W) \displaystyle \sum _{i=1}^{8} h_i \eta^{a_i}.
\end{equation}
With $\eta = \alpha_s(M_W) /\alpha_s (m_b)$,
$$ h_i =\left( \frac{626126}{272277}, -\frac{56281}{51730},
-\frac{3}{7}, -\frac{1}{14}, -0.6494, -0.0380, -0.0186, -0.0057 \right),$$
$$a_i = \left( \frac{14}{23}, \frac{16}{23}, \frac{6}{23}, -\frac{12}{23},
0.4086, -0.4230, -0.8994, 0.1456 \right).$$

\section{The $B \rightarrow X_s \gamma$ decay rate and phenomenology}

Following refs.\cite{Ciu}, applying a spectator model,
\begin{eqnarray}
BR(B \rightarrow X_s \gamma) /BR(B\rightarrow X_c e\overline{\nu})
\simeq\Gamma(b\rightarrow s\gamma)/\Gamma
(b\rightarrow ce\overline{\nu}).
\end{eqnarray}
Then when have
\begin{eqnarray}
\frac{BR(B \rightarrow X_s \gamma)}{BR(B
\rightarrow X_c e \overline{\nu})} \simeq \frac{|V_{tb} V_{ts}^*|^2}
{|V_{cb}|^2} \frac{6 \alpha_{QED}}{\pi
 g (m_c/m_b)}
|C_7^{eff}(m_b)|^2 \left(1-\frac{2 \alpha_{s}(m_b)}{3 \pi} f(m_c/m_b)
\right)^{-1},
\end{eqnarray}
where the phase space factor $g(z)$ is given by:
\begin{eqnarray}
g(z)=1-8z^2+8z^6-z^8-24z^4\log z,
\end{eqnarray}
and the factor $f(m_c/m_b)$ of one-loop QCD correction to the semileptonic
decay is,
\begin{eqnarray}
f(m_c/m_b)=(\pi^2 -31/4)(1-m_c^2/m_b^2) + 3/2.
\end{eqnarray}
 Afterwards one obtains the $B \rightarrow X_s
\gamma$ decay rate normalized to the quite well established
 semileptonic decay rate $Br(B \to X_c e\overline{\nu} )$.
If we take experimental result $BR(B \to
X_c e\overline{\nu} ) =10.8\% $\cite{data}, the branching ratios of
$B \to X_s \gamma$ is found to be:
\begin{eqnarray}
BR(B \rightarrow X_s \gamma)
\simeq 10.8\%\times \frac{|V_{tb} V_{ts}^*|^2}
{|V_{cb}|^2} \frac{6 \alpha_{QED}\;|C_7^{eff}(m_b)|^2}
{\pi g (m_c/m_b)}
 \left(1-\frac{2 \alpha_{s}(m_b)}{3 \pi} f(m_c/m_b)
\right)^{-1}.
\end{eqnarray}
In numerical calculations we always use
$M_W=80.22\;GeV$, $\alpha_s(m_Z)=0.117$,
$m_c=1.5\;GeV$,  $m_b=4.8\;GeV$ and  $|V_{tb} V_{ts}^*|^2/
|V_{cb}|^2= 0.95$ \cite{data} as input parameters.

Generally speaking, the contribution to the decay $b\to s \gamma$ from
color singlet $P^\pm$ is small
when compared with the contribution from the color octet $P_8^\pm$,
since there is a color enhancement factor $8$ appeared in the third terms
in eqs.(\ref{c7}, \ref{c8}) for the functions $B(y)$ and $D(y)$.
 Fig.2 is the plot of the branching ratio $Br(B \rightarrow X_s \gamma)$
as a function of the top quark mass.
The upper dashed curve in Fig.2 represents
the branching ratio in the standard model, while the solid curve shows
the same ratio with the inclusion of the contributions from
$P^{\pm}$ and $P_8^{\pm}$ assuming $m(P^\pm) =300\;GeV$ and
$m(P^\pm_8) =600\;GeV$. The band between two dash-dotted lines
corresponds to the newest CLEO limits: $ 1.0\times 10^{-4} <
BR(B \to X_s \gamma) < 4.2\times 10^{-4}$ at $95\%C.L.$ \cite{cleo2}.
The branching ratio of $b \to s\gamma$ with large contribution
from OGTM, is much more sensitive with the top quark mass, compared with
the case of pure SM.

It is known from the decoupling theorem that for heavy enough
nonstandard boson, we should recover the SM result. So for
sufficiently large values of $m(P^{\pm})$, $m(P^{\pm}_8)$ (
e.g. $m(P^{\pm})>600$GeV, $m(P^{\pm}_8)>2000$GeV), the
contributions from OGTM shall be negligible. This can also be seen
from the fact that the functions $B$, $D$ and $E$ go to zero, as
$x$,$y\to \infty$.  For
not so large $m(P^{\pm})$, $m(P^{\pm}_8)$, the OGTM contribution cancels much
of the SM contribution because of the relative minus sign between
their contribution. As a result, the branching of $b\to s\gamma$
reached the lower limit of the CLEO experiment. So a large region of
$m(P^{\pm})$, $m(P^{\pm}_8)$ ( i.e. $1000GeV <m(P^{\pm}_8)<2000GeV$,
for all
$m(P^{\pm})$ ) is ruled out. When $m(P^{\pm})$, $m(P^{\pm}_8)$ go on
smaller, their contribution is about two times as large as contribution
of SM (recall there is a relative minus sign), the branching ratio
of $b \to s\gamma$ resumes to experiment allowed region. But if the
$m(P^{\pm})$, $m(P^{\pm}_8)$ are smaller enough, the contribution
of OGTM is more larger, the region is also excluded by the upper limit
of CLEO experiment. The whole result is
 illustrated at Fig.3, large part of the parameter space in the
$m(P^\pm)-m(P_8^\pm)$ plane can be excluded according to the current $CLEO$
95\% C.L. limits on the ratio $BR(B \rightarrow X_s \gamma)$ \cite{cleo2}.
 It is easy
to see that no direct limits on $m(P^\pm)$ can be obtained at present
for free $m(P^{\pm}_8)$,
but at the same time, one can simply read out the lower
bound on the mass of color octet PGBs:
$m(P^{\pm}_8) > 440\;GeV$ for free $m(P^{\pm})$
(assuming $m_t=180\;GeV$), if we simply interpret the $CLEO$
 $95\% C.L.$ limits  on the ratio  $BR(B \rightarrow X_s \gamma)$
as the bounds on the masses of charged PGBs.

Of cause, we have not considered the effects of other possible
uncertainties, such as that  of $\alpha_s(m_Z)$, next-to-leading-log
QCD contribution\cite{AJB}, QCD correction from $m_{top}$ to $M_W$\cite{lcd}
 etc. The inclusion of those
additional uncertainties will broaden the border lines between the
allowed regions and excluded regions in Fig.3. The limitations drawn
from the calculations will be surely weaken, i.e., the lower
limit will become $m(P^{\pm}_8) > 400\; GeV$ at $90\%C.L.$ if we include
an additional $20\%$ theoretical uncertainties.

As a conclusion, the size of contribution to the rare decay of $b\to s \gamma$
from the PGBs strongly depends
on the values of the masses of the top quark and the charged PGBs. This is
quite different from the SM case. By the comparison
of the theoretical prediction with the current data one can derives out
the constraints on the masses of the color octet charged PGBs:
$m(P^\pm_8) > 400\;GeV$ at $90\%C.L.$ for free $m(P^\pm)$, assuming
$m_t=180\;GeV$.

\vspace{1cm}
\noindent {\bf ACKNOWLEDGMENT}

This work was supported in part by the National Natural Science
Foundation of China, the funds from
Henan Science and Technology  Committee and China Postdoctoral Science
Foundation.


\begin{center}
{\bf Figure Captions}
\end{center}
\begin{description}

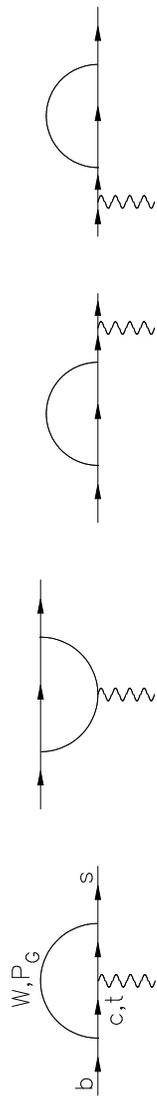
\item[Fig.1:] The Feynman diagrams which contribute to the rare radiative
decay $b \to s \gamma$. The half-circle-lines in the loop represent the
W gauge boson and charged PGBs propagators.

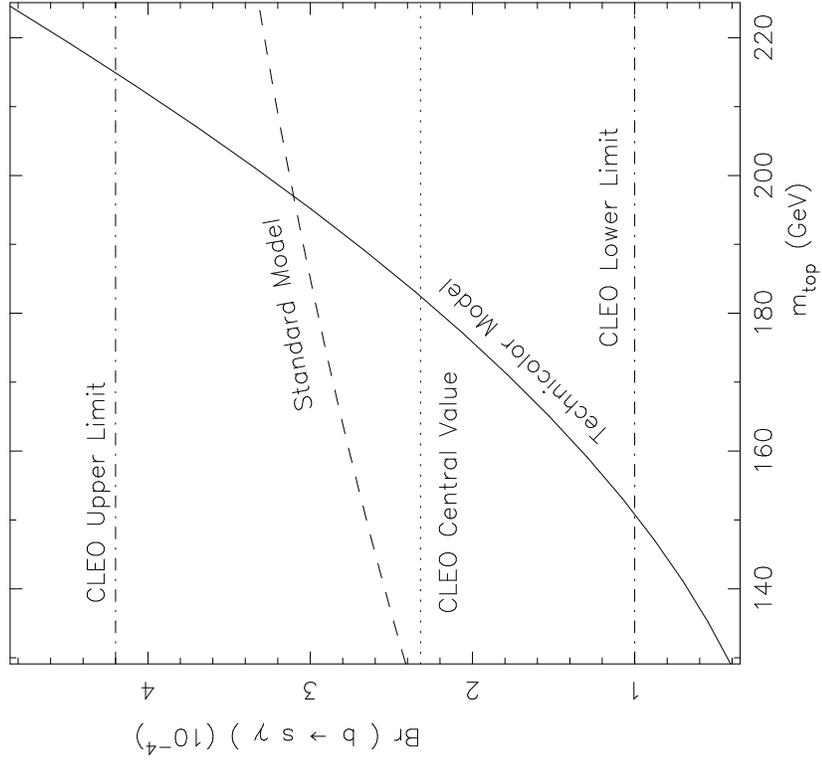
\item[Fig.2:] The plot of the branching ratio of $b\to s \gamma$ versus the top
quark
mass $m_t$ assuming $m(P^{\pm})=300\;GeV$ and $m(P^{\pm}_8)=600\;GeV$.
For more details see the text.

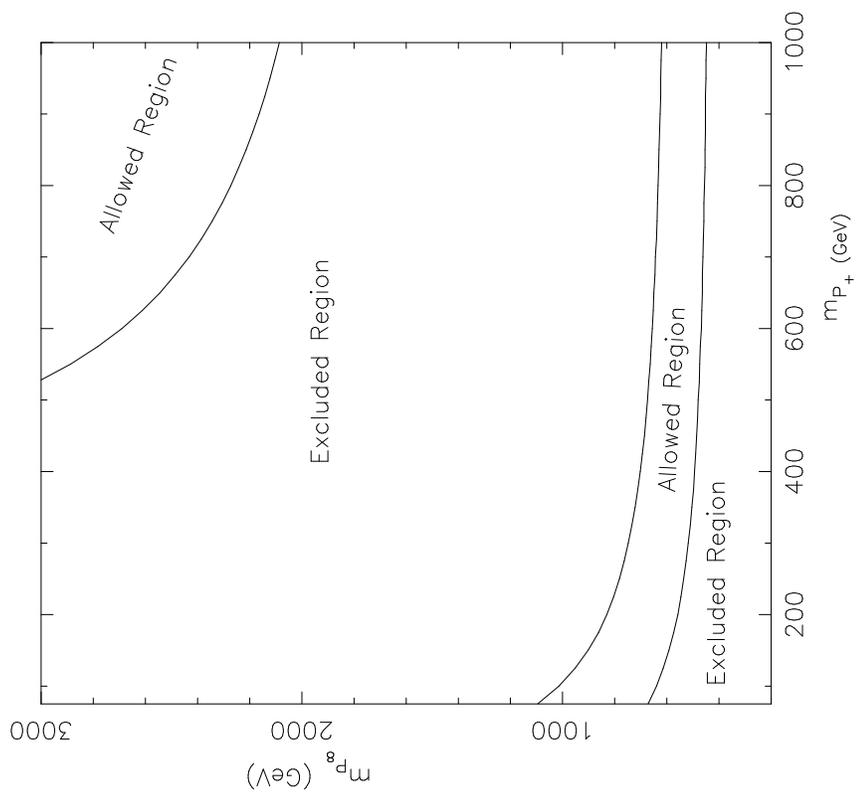
\item[Fig.3:] Allowed range in the $m(P^\pm)-m(P_8^\pm)$ plan for
$m_t=180\;GeV$, the band is corresponding to
the current $CLEO$ 95\% C.L. limits on the ratio $BR(B \to X_s \gamma)$
as given in eq.(2).

\end{description}

\end{document}